\documentstyle[12pt]{article}
\renewcommand{\(}{\left(} \renewcommand{\)}{\right)}
\renewcommand{\[}{\left[} 

\newcommand\ba{\begin{array}}
\newcommand\ea{\end{array}}
\newcommand\ben{\begin{equation}}
\newcommand\een{\end{equation}}
\newcommand\bea{\begin{eqnarray}}
\newcommand\eea{\end{eqnarray}}
\newcommand{\al}{\alpha}
\newcommand{\be}{\beta}
\newcommand{\ga}{\gamma}
\newcommand{\de}{\delta}

\newcommand{\rf}[5]{#1, {\it #3} {\bf #4}, #5 (#2)}

\newcommand{\MPL}{Mod. Phys. Lett.}
\newcommand{\RPP}{Rep. Prog. Phys. }

\newcommand{\PL}{Phys.\ Lett.\ }

\newcommand{\RMP}{Rev.\ Mod.\ Phys.\ }

\newcommand{\R}{R}

\newcommand{\bx}{\mbox{\boldmath $x$}}
\newcommand{\pa}{\partial}
\newcommand{\tr}{{\rm tr}}

\title{
{\normalsize \begin{flushright}SUSX--HEP--94--73\\NI94028\\
{\tt cond-mat/9501026}\\
December 1994\\
\end{flushright}}
\vspace{2 cm}
Where are the Hedgehogs in Nematics?}
\author{Mark Hindmarsh\thanks{E-mail: {\tt m.b.hindmarsh@sussex.ac.uk}}}
\date{}

\begin{document}
\maketitle
\vspace{-24pt}
\begin{center}
{\normalsize {\it
School of Mathematical and Physical Sciences\\
University of Sussex\\
Brighton BN1 9QH\\
UK\\[6pt]
{\rm and}\\[6pt]
\it Isaac Newton Institute for Mathematical Sciences\\
20 Clarkson Road\\
Cambridge CB3 0EH\\
UK}\\[12pt]
{\sc PACS} numbers: 61.30.Jf 64.70.Md
}
\end{center}

\vfill

\begin{abstract}

In experiments which take a liquid crystal rapidly from the
isotropic to the nematic phase, a dense tangle of defects is formed.
In nematics, there are in principle both line and point defects
(``hedgehogs''), but no point defects are observed until the
defect network has coarsened appreciably.  In this letter the
expected density of point defects is shown to be extremely low,
approximately $10^{-8}$ per initially correlated domain, as result of the
topology (specifically, the homology) of the order parameter space.

\end{abstract}
\newpage

An outstanding puzzle in the formation of defects after rapid quenches
is the absence of point disclinations (``hedgehogs'' or ``monopoles'')
in nematic liquid crystals.  One might naively expect of order one
defect per initially correlated domain.  However, experiments with
rapid pressure quenches \cite{Chu+91} do not find any monopoles at all
to begin with, although the expected dense network of line
disclinations (``strings'') is present, and rapidly reaches a scaling
regime in which the string length density decreases as $t^{-1}$.  This
can be understood to be a result of the $t^{1/2}$ growth law of the
network scale length $L(t)$. If there is on average about 1 segment of
string of length $L$ per volume $L^3$, the scaling of the length
density $L^{-2}$ is explained.  The monopoles do not make their
appearance until relatively late in the scaling regime, forming from
collapsing loops of string.

In this letter a solution to this puzzle is presented.
The answer lies in the topology of
the order parameter space, which is the projective plane $\R P^2$.  In
order for there to be a monopole inside some sphere in the liquid
crystal, the order parameter field has to cover its entire space
twice.  It turns out that this is very hard to arrange out of the random
initial conditions produced by a rapid quench. There is an
underlying mathematical formulation of the solution, in terms of the
homology of the order parameter space, which is outlined briefly at the
end of this work.

The order parameter of a liquid crystal is a traceless symmetric rank 2
tensor $Q_{ij}(\bx)$.  The normalized eigenvector with the largest
eigenvalue is known as the director field $n_i(\bx)$, for it defines
the average local orientation of the liquid crystal molecules.  In a
nematic the other two eigenvalues are equal, and we may write
\cite{NLC}
\ben
Q_{ij} = A(n_in_j - \frac{1}{3}\de_{ij}).
\een
The free energy of the liquid crystal is
\ben
F[Q] = \int d^3x \( L_1\pa_kQ_{ij} \pa_kQ_{ij} + L_2 \pa_jQ_{ij} \pa_kQ_{ik}
+ L_3\pa_kQ_{ij}\pa_jQ_{ik}+V(Q)\)
\een
where $V(Q)$ is the bulk free energy.  Near the phase transition we are
justified in expanding to quartic order, and
\ben
V(Q) = \frac{1}{2} \al\, \tr(Q^2) + \frac{1}{3} \beta\, \tr(Q^3) + \frac{1}{4}
\ga \,\tr(Q^2)^2 + \cdots.
\een
The condition that the system be in the nematic phase, i.e.~that $A\ne
0$ minimize the free energy, is just $\al\ga/\be^2 \le 1/9$.
In this phase the symmetry group of the bulk free
energy density, which is the group of spatial rotations SO(3), is
reduced to the cylinder group $D_{\infty}$, or O(2).  The manifold $M$
of possible equilibrium states is defined by the condition
 $\de F /\de Q = 0$, subject
to the constraints of tracelessness and symmetry.  This is isomorphic
to the coset space SO(3)/O(2), or the real projective plane $\R P^2$, which
can be thought of as a 2-sphere with antipodal points identified.

After a rapid quench, the order parameter is uncorrelated beyond a
certain distance $\xi_0$, which is determined by the relative
magnitudes of
the quench time and the relaxation time of the system.  The
isotropic-nematic transition is weakly first order, which means that
the correlation length grows by a large factor as the phase transition
is approached \cite{NLC}, although the transition itself appears to proceed by
bubble nucleation and growth \cite{Bow+94}.  The substance of the
present argument should not depend on the order of the transition: it is
only the existence of uncorrelated domains that is important.  Defects
form at the interstices of domains where the order parameter is in some
sense maximally misaligned, in what is known in cosmology as the Kibble
mechanism \cite{Kib76}. One can estimate the density of defects by
applying the so-called ``geodesic rule'' \cite{Vac91,RudSri,Hin+94}.  This
assumes that if we pick two points $\bx_1$ and $\bx_2$ in adjacent
domains with order parameters $Q^1$ and $Q^2$, the most likely
interpolation $Q^{12}(\bx)$ on a line
between the points is the shortest path in
$M$, for this locally minimizes the bulk free energy.
Since $M$ comes equipped
with a metric by virtue of its embedding in the Euclidean field space,
this path is by definition a geodesic.     Now consider three adjacent
domains, and pick three points $\{\bx_1,\bx_2,\bx_3\}$.  The geodesic
rule can be applied separately to each pair of domains, and then to all
three: the interpolation $Q^{123}(\bx)$ to the interior of the triangle
$\{\bx_1,\bx_2,\bx_3\}$ is a geodesic {\em surface} in $M$.  There may
be an obstruction to the procedure:  if the loop
$\{Q^{12},Q^{23},Q^{31}\}$ is in the non-trivial homotopy class of
$\pi_1(M)$, a
line defect must pass through the triangle $\{\bx_1,\bx_2,\bx_3\}$, at
the junction of the three correlated domains.  A similar argument
involving four domains is applied to the formation of point defects
\cite{Kib76},
which are obstructions to the construction of an interpolating geodesic
3-simplex $Q^{1234}$. We shall see, however, that four uncorrelated
domains are not enough for a point disclination in a nematic liquid
crystal.

Calculating the probability of finding a defect associated with
non-trivial $\pi_n(M)$  at the
interstices of $n+1$ domains is a problem in geometric probability on
the manifold of equilibrium states $M$.  This problem has been solved
only for $M\simeq S^n$  \cite{LeePro91}, and
for 1-dimensional defects in $\R P^2$ \cite{Vac91} and $S^3/Z_2$
\cite{KibZ2}. The solution is rather neat for the spheres.  Consider
first $n=1$, where the order parameter is a 2 component field $\phi_a$
with $\sum_a\phi_a^2$ constant. The problem consists essentially of placing 3
points $\phi^1$, $\phi^2$, $\phi^3$ at random on the circle of constant
$\sum_a\phi_a^2$, and asking the probability for $\phi^3$ to lie between
$-\phi^1$ and $-\phi^2$ (taking the shortest route).  In that case, and
in only that case, will the geodesic rule supply a loop which wraps
around $M$.  Now, $\pm \phi^1$ and $\pm\phi^2$ divide the circle into
4. Given that $\phi^1$ and $\phi^2$ are isotropically distributed, one
can convince oneself that the average length of the line segment
between  $-\phi^1$ and $-\phi^2$ is 1/4.  This is then the  probability
of finding a line defect at the junction of three adjacent domains,
and the number of defects per unit area is therefore $1/4\xi_0^2$.  This
generalizes for arbitrary $n$ to $1/2^{n+1}$.  For strings in $\R P^2$ the
calculation is more involved, but it emerges that the
probability is $1/\pi$.

The problem with trying to extend these calculations to point defects
in $\R P^2$ is that four neighbouring uncorrelated domains can never
generate such a defect.  To construct a hedgehog configuration of the
order parameter we must cover $M$ twice, because the director field has
an $\bx \to -\bx$ symmetry.  One cannot unambiguously do this with four
domains, for the geodesic rule produces a mapping from the tetrahedron
$\{\bx_1,\bx_2,\bx_3,\bx_4\}$ which is either trivial or contains a
string passing through two of the faces.  The point is that in order
to cover $M$ twice, each face of the tetrahedron must cover on average
half of it, which means that there will always be faces trying to
cover more than half. This cannot happen with the geodesic rule.  Thus
we need more domains, which inevitably lowers the probability of
finding a defect.

The minimal triangulation of $\R P^2$ has in fact 6 vertices (see Figure
1).  One can think of this as a triangulation of $S^2$ by an
icosahedron, with antipodal points then identified.  Thus in order to
cover $\R P^2$ twice we need a roughly spherical arrangement of a
minimum of 12 uncorrelated adjacent domains.  A great deal of
calculation can now be saved by an approximation which uses a fixed
triangulation of $\R P^2$ directly.   For example \cite{VacVil84,HinStr94},
if we
approximate $S^1$ by three equidistant points labelled 0,1,2, and
assign a string to a spatial triangle $\{\bx_1,\bx_2,\bx_3\}$ when all
three values of $\phi$ are different, the probability of having a
string passing through the triangle is  just the number of different
arrangements of 0,1, and 2 divided by the total number of possible
assignments $3^3$. Thus the probability in this discrete approximation
of 1-dimensional defect passing through the triangle is
\ben
P'_1(S^1) = 3!/3^3 = 2/9, \een
where the prime is used to denote the approximation to the true
geometric probabilities $P_n(M)$.  For general $n$ we have
\ben
P'_n(S^n) = (n+2)!/(n+2)^{n+2}.
\een
This approximation gets worse for large $n$.
Using Stirling's approximation, one sees that
$P'_n(S^n)/P_n(S^n)\sim n^{1/2} e^{(\ln 2 -1)n}$.

For line defects in $\R P^2$ the calculation proceeds as follows.  The
first two values of the order parameter $Q^1$ and $Q^2$ can be any two
different vertices of the triangulation.  The last point must be one of
the two which are connected to both of the first two.  Thus
\ben
P'_1(\R P^2) = 6\cdot 5\cdot 2/6^3 = 10/36
\een
which is close to the true value $P_1(\R P^2) = 1/\pi$.
For point defects, we must calculate the number of different ways of
assigning values of $Q$ to the 12 domains. Picking any two adjacent domains,
the first values can once again be any two vertices of the triangulation.
In a third domain, adjacent to both the first two, one must
correspondingly pick one of the vertices connected to both those
already selected. Thus
\ben
P'_2(\R P^2) = 60/6^{12} \simeq  2.76 \times 10^{-8}.
\een
A quick way of calculating this number is to note that the
assignment of vertices to domains is just a map from one
icosahedron to another with opposite points identified. Therefore
$P'_2(\R P^2)$ is just the order of the icosahedral group, which is
120,  divided by 2.

The configuration of
domains occupies a volume of
approximately $\xi_0^3$, and so the density of point defects $N_p$
is roughly
\ben
N_p \simeq 10^{-8} \xi_0^{-3}.
\een
This is a very small number, as promised.
If the discrete calculation is here as good an approximation as for the
spheres, then it explains why the point defects of a nematic liquid
crystal are not found after a rapid quench: they require a very special
arrangement of the order parameter over many uncorrelated domains
\cite{MZNB}.

The icosahedral arrangement
of domains can be extended into the body of the
material  by the addition of a further domain in the
centre.  One then realises that the ``point'' disclination is
actually a small loop of size $\sim \xi_0$ encircling the
central domain.  The value of the order parameter here merely controls
the loop's orientation.  Thus there is a sense in which there are {\em
no} point disclinations at all.  What we have calculated is merely
the density of the smallest possible loops which can form hedgehogs.

To conclude, I outline the mathematical structure implicit in the
geodesic rule.  Recall that the construction starts with
points $\{\bx_i\}$ in uncorrelated domains, and the corresponding
values of the order parameter $\{Q^i\}$.  One attempts to construct
an approximation to the field configuration over the whole of $\R^3$
by extending the points
$\{\bx_i\}$ to a full triangulation, defining the order
parameter field $Q(\bx)$ by the geodesic rule. This determines how
to ``fill in'' the set of closed figures (points, lines, and triangles)
in order to create others
(lines, triangles and tetrahedra) of higher dimension.
The result is a simplicial complex \cite{Homol}
in the order parameter space $M$. However, the procedure
fails when some sub-complex cannot be filled in, that is,
the sub-complex is not the boundary of another, higher-dimensional
complex in the space $M$. The order parameter has to leave $M$ and
a defect appears in the corresponding region of $\R^3$. This
can happen if and only if the space has a non-trivial homology group
$H_n(M)$.  Thus the Kibble mechanism coupled with the geodesic rule
produces defects of dimension $d$ in a space of dimension $D$ only if
$H_n(M)$, with $n=D-d-1$, is non-trivial.  The second homology class
of $\R P^2$ is zero: this is the underlying reason for the absence of
point defects in nematic liquid crystals.

I am indebted to Neil Turok for many useful discussions.  I also thank
Martin Zapotocky for encouragement to write this work up, and for a
careful reading of the first draft of the manuscript.
I also wish to acknowledge the hospitality of the Isaac
Newton Institute where this paper was completed.
This work is supported by PPARC Advanced Research Fellowship B/93/AF/1642.

\newpage

\section*{Figure Captions}

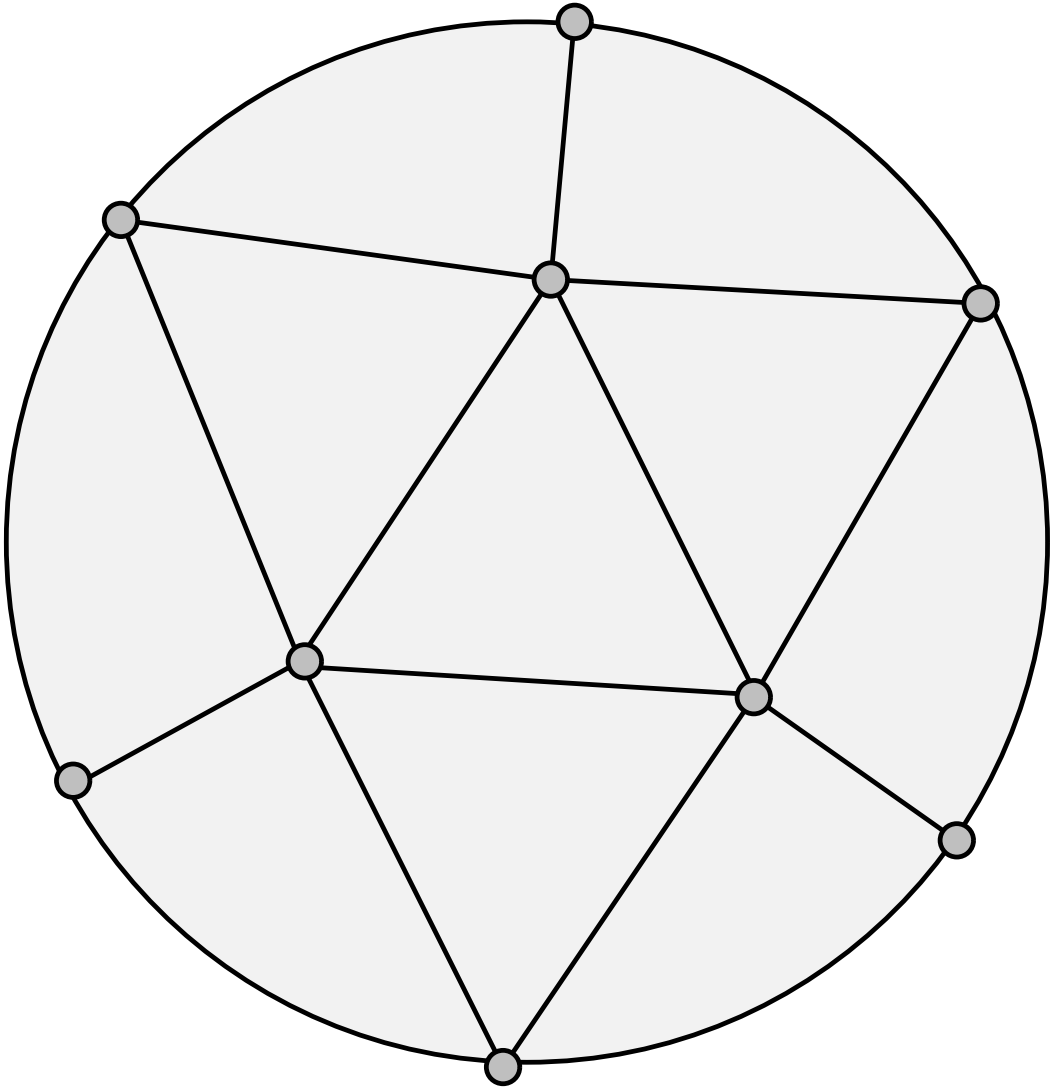
{\bf Figure 1:} The minimal triangulation of $\R P^2$, consisting
of 6 vertices, 12 edges and 10 faces. This is essentially the top half
of an icosahedron.  Opposite points on the boundary are identified.


\begin{thebibliography}{99}
\bibitem{MZNB} It has been pointed
out to me by Martin Zapotocky that the same calculation should also apply
to $\R P^2$ textures in 2 dimensions.
\end{thebibliography}
\end{document}